\documentclass[12pt,preprint2]{aastex}
\usepackage{emulateapj5}
\usepackage{onecolfloat5}
\usepackage{epsf}

\begin{document}

\shorttitle{Lensing Probability of High Redshift Quasars}
\shortauthors{Comerford, Haiman \& Schaye}
\slugcomment{Submitted to The Astrophysical Journal}

\twocolumn[
\title{Constraining the Redshift  $z\sim6$ Quasar \\ Luminosity 
Function Using Gravitational Lensing}
\author{Julia M. Comerford, Zolt\'an Haiman\altaffilmark{1}} 
\affil{Princeton University Observatory, Princeton, NJ 08544, USA}
\email{comerfrd@astro.princeton.edu, zoltan@astro.princeton.edu}
\author{Joop Schaye} 
\affil{Institute for Advanced Study, School of Natural Sciences, 
Olden Lane, Princeton, NJ 08540, USA}
\email{schaye@ias.edu}

\begin{abstract}
Recent discoveries by the Sloan Digital Sky Survey (SDSS) of four bright $z
\sim 6$ quasars could constrain the mechanism by which the supermassive black
holes powering these sources are assembled. Here we compute the probability
that the fluxes of the quasars are strongly amplified by gravitational lensing,
and therefore the likelihood that the black hole masses are overestimated when
they are inferred assuming Eddington luminosities. The poorly--constrained
shape of the intrinsic quasar luminosity function (LF) at redshift $\sim 6$
results in a large range of possible lensing probabilities.  If the LF is
either steep, or extends to faint magnitudes, the probability for amplification
by a factor $\mu\ga10$ (and with only one image detectable by SDSS) can
reach essentially $100 \%$.  We show that future observations, in particular,
either of the current four quasars at the high angular resolution provided by
the Hubble Space Telescope, or of an increased sample of $\sim 20$ redshift six
quasars at the current angular resolution, should either discover several
gravitational lenses, or else provide interesting new constraints on the shape
of the $z\sim 6$ quasar LF.
\end{abstract}

]
\altaffiltext{1}{Hubble Fellow}

\section{Introduction}
\label{sec:intro}
The Sloan Digital Sky Survey (SDSS) has recently discovered several extremely
bright and distant quasars with redshifts $z \sim 6$, whose luminosities are a
result of gas accretion onto supermassive black holes (BHs). These rare objects
have luminosities at the far bright tail of the quasar luminosity function
(LF), corresponding to $\sim 5 \; \sigma$ peaks in the density field in models
in which quasars populate dark matter halos (Haiman \& Loeb 2001; Fan et
al. 2001b). By assuming that the observed luminosities equal the limiting
Eddington luminosities of the BHs, the masses of all these BHs are inferred to
be a few $\times 10^9~{\rm M_\odot}$.  The mere existence of such massive BHs
at so early a stage in the evolution of the universe should provide insight
into the formation and growth of supermassive black holes (Haiman \& Loeb
2001).

The above estimates for the masses of the BHs associated with the SDSS quasars
scale directly with the inferred luminosities, and thus with the observed
fluxes of the quasars. However, the apparent fluxes can be increased by
gravitational lensing. If a quasar is gravitationally lensed and this is not
taken into account in the mass estimate, the quasar's intrinsic luminosity, and
therefore its BH mass, will be overestimated. Previous considerations of the
probability that these SDSS quasars are lensed indicate that while the a~priori
lensing probability is small, the magnification bias is poorly known and can be
quite large (Wyithe \& Loeb 2002).  If massive BHs grow from stellar--mass
seeds by accreting gas at the Eddington rate, they reach masses of $\sim
10^9~{\rm M_\odot}$ in $\sim 10^9$ years, a time span comparable to the age of
the universe at $z \sim 6$ in the current ``best--fit'' cosmologies.

Since the presence of such massive BHs can only be marginally accommodated into
structure formation models (Haiman \& Loeb 2001), it is interesting to ask
whether {\it{all}} of the SDSS quasars may be strongly lensed.  In this paper,
we compute the expected probabilities for lensing of $z\sim6$ quasars by
intervening galaxies. We quantify how the probability depends on the shape of
the assumed quasar LF, and in particular, we focus on the question: can the
expected probabilities reach values near unity? We show that current
constraints on the $z\sim 6$ quasar LF are sufficiently weak that the
probabilities for strong lensing can indeed be close to $100\%$. We will
utilize this result to ``invert`` the problem, and demonstrate how the
intrinsic LF can be constrained using current and future detections (or
absence) of lensing events with multiple images.

The rest of this paper is organized as follows. In \S~2, we describe our model
for the population of gravitational lenses. In \S~3, we derive the resulting
intrinsic ``a~priori'' lensing probabilities along a random line of sight to
redshift $z\sim 6$.  In \S~4, we discuss the effects of magnification bias, and
the expected a~posteriori lensing probabilities for the $z\sim 6$ quasars in
models with different LFs.  In \S~5, we quantify constraints on the quasar LF
that can be derived from current and future searches for lensing
events. Finally, in \S~6, we discuss our main results and then summarize the
implications of this work.  Throughout this paper, we adopt a flat cosmological
model dominated by cold dark matter (CDM) and a cosmological constant
($\Lambda$), with $\Omega_m=0.25$, $\Omega_b=0.04$, and $\Omega_\Lambda=0.75$,
a Hubble constant $H_0=70~{\rm km~s^{-1}}$, an rms mass fluctuation within
a sphere of radius $8 \; h^{-1}$ Mpc of $\sigma_8=0.9$, and power--law index
$n=1$ for the power spectrum of density fluctuations. We also adopt the
cosmological transfer function from Eisenstein \& Hu (1999).

\section{Model description}
\label{sec:model}

The a~priori probability of gravitational lensing depends primarily on the
comoving number density of lenses and the lensing cross section of each
lens. Present theories agree that galaxies are created by the condensation of
gas in dark matter halo cores, and the abundance of the most massive halos
depends on the overall amplitude of mass fluctuations. We use the halo mass
function of Jenkins et al.\ (2001), who measure $dn/dM$, the number of halos
per unit comoving volume and per unit mass, from $N$-body simulations.

We employ a combination of two separate lens density profiles. The simpler of
the two halo density profiles we consider is the singular isothermal sphere
(SIS) density profile. The SIS profile is characterized by its dependence on
the one-dimensional velocity dispersion, which is supported by observations of
flat rotation curves, and its density profile is $\rho(r)=\sigma_v^2/2 \pi G
r^2$, where $\sigma_v$ is the velocity dispersion. The steep inner density
slope of the SIS profile, however, renders it a poor fit to the more massive,
and more extended halos. As a result, Navarro, Frenk, \& White (1997; hereafter
NFW) proposed a ``universal'' mass density profile for dark matter halos found
in numerical simulations. The NFW density profile is $\rho(r)=\rho_s r_s^3 /
r(r+r_s)^2$, where $\rho_s$ and $r_s$ are constants. Although it matches the
SIS model at intermediate radii, it branches off for large and small radii.

Because of the dynamics of cooling gas in halos (Keeton 1998; Porciani \& Madau
2000; Kochanek \& White 2001; Li \& Ostriker 2002), we use a combination of the
SIS and NFW profiles to describe the halos that contribute to the quasar
lensing probability. Gas falling into very massive halos is shock--heated to
high temperatures ($T\ga 10^7$K) and cannot cool within a Hubble time. The gas
therefore remains in hot, extended gaseous halos around groups or clusters of
galaxies which have shallow inner density profiles. For less massive halos, in
contrast, the virial temperature is low enough that the gas can cool and
collapse into galaxies with steep inner density slopes. This difference in
density slopes implies that a single density profile cannot describe all
lensing objects in the universe.

As a solution, we use the SIS profile ($d\rho/dr \sim r^{-3}$) for less massive
halos which have steep inner density slopes and the NFW profile ($d\rho/dr \sim
r^{-2}$ for $r \ll r_s$) for more massive halos which have shallow inner
density slopes. We set the transition between these two profiles at a dark
matter halo mass of $10^{13} {\rm M_\odot}$, the mass at which typically half
of the dark halos have cooled the majority of their gas content, as given by
the halo cooling probability distribution in Kochanek \& White (2001).  While
this value of the cutoff has a physical basis, perhaps more importantly the
predicted splitting angle distribution for lensing events with multiple images
matches observations (Keeton 1998; Kochanek \& White 2001; Li \& Ostriker
2002). Therefore, in our calculations, we used the SIS profile for halo masses
less than $10^{13}~{\rm M_\odot}$ and the NFW profile for halo masses greater
than $10^{13}~{\rm M_\odot}$.

NFW halos are very inefficient lenses, with a cross--section per unit mass $\ga
100$ times smaller than for SIS halos.  Although the high mass halos are
responsible for most lensing events with large ($\ga 5^{\prime\prime}$) image
splittings, these comprise only a small ($\sim 1\%$) fraction of all lensing
events of interest. As a result, in practice, we find that we can safely ignore
NFW halos, and use a model where halos with masses below $10^{13}~{\rm
M_\odot}$ are assumed to have SIS profiles, and halos with masses above
$10^{13}~{\rm M_\odot}$ are simply excluded from the calculations.  Our results
for the intrinsic lensing probability are relatively insensitive to the choice
of this mass cutoff: a factor of two higher/lower value would increase/decrease
the intrinsic lensing probability to $z\sim 6$ by $\approx 20\%$. The effect on
the a~posteriori lensing probability is similar in models where the probability
is relatively low ($\la 50\%$), but the mass cutoff has a smaller effect on the
a~posteriori lensing probability if this probability is high (see discussion
below).

\begin{figure}[ht!]
\begin{center}
\plotone{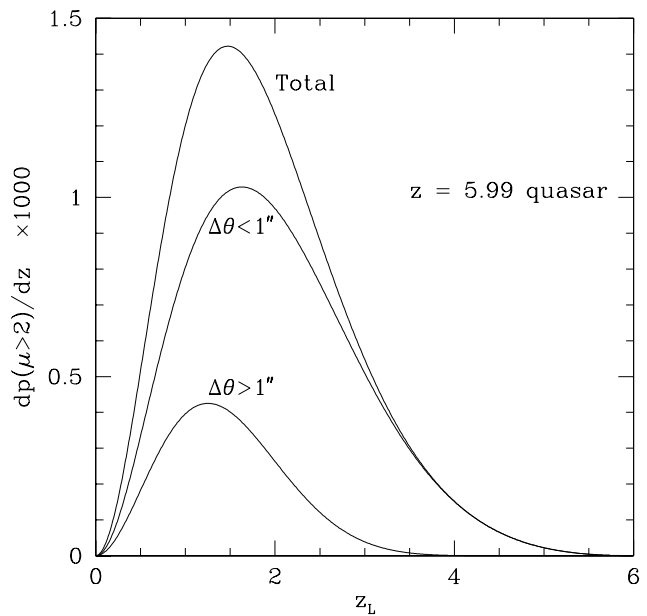}
\caption{The a~priori probability as a function of lens redshift that the
$z=5.99$ SDSS quasar was magnified by at least a factor $\mu_{\rm min}=2$,
but has not produced two separately detectable images for SDSS.  The middle
curve in the figure shows the probability for multiple images with an image
splitting of $\Delta\theta<1^{\prime\prime}$.  The bottom curve shows the
probability for image splittings of $\Delta\theta>1^{\prime\prime}$, but
subject to the constraint that the fainter image must be below the SDSS
detection limit (this corresponds to $2\leq\mu\leq6.172$ for this source).  The
top curve shows the total lensing probability.}
\label{fig:dtaudz}
\end{center}
\end{figure}

\begin{figure}[ht!]
\begin{center}
\plotone{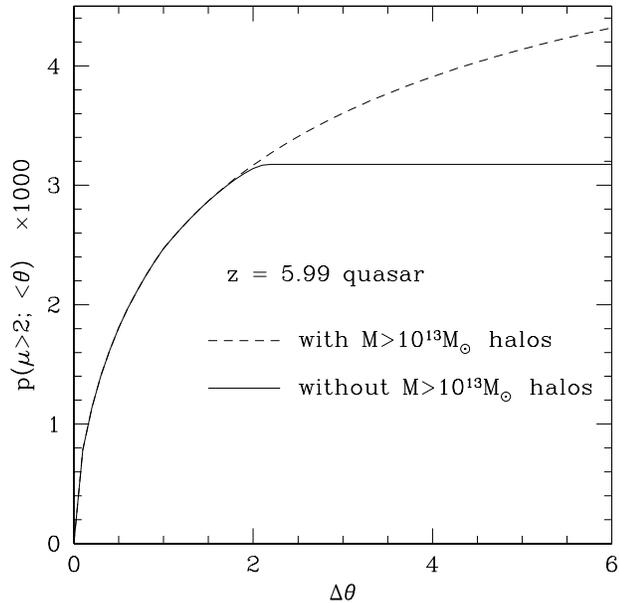}
\caption{The a~priori probability for the $z=5.99$ SDSS quasar to be magnified
by at least a factor of two, as a function of the maximum allowed splitting
angle $\Delta\theta$. The dashed curve includes all halos, and assumes that
each halo is described by an SIS profile. The solid curve excludes halos above
$10^{13}{\rm M_\odot}$. Including halos above this mass, described by NFW
profiles, would yield a curve that is nearly identical to the solid curve. For
splitting angles of $\Delta\theta>1^{\prime\prime}$ we require that the fainter
image must be below the SDSS detection threshold. This excludes lensing events
with $\mu\geq6.172$, and causes a $\sim10\%$ decrease in the lensing
probability for $\Delta\theta>1^{\prime\prime}$.}
\label{fig:dtaudtheta}
\end{center}
\end{figure}

\section{A~priori lensing probability}
\label{sec:apriori}
The {\it{a~priori}} probability that a quasar is lensed by an intervening
galaxy is given by the product of the number of lenses per unit volume,
$(1+z)^3 \; dn/dM$, the galaxy lensing cross section, $\sigma_L$, and the path
length of light from the quasar, $c \; (dt/dz) \; dz$, where
$dt/dz=-1/(1+z)H(z)$, integrated over all possible lens masses $M_L$ and
redshifts $z_L$.  Therefore, the a~priori probability that a source at redshift
$z_s$ is lensed and magnified by a factor $\geq \mu$ is
\begin{eqnarray}
\nonumber
p(\mu, z_s)&=& \int_0^{\infty} \int_0^{z_s} c \frac{(1+z_L)^2} {H(z_L)}
\sigma_L(>\mu; M_L, z_L, z_s)\\
&& \times \frac{d n} {d M} (M_L, z_L) dz_L dM_L ,
\label{eq:lensdepth}
\end{eqnarray}
where the Hubble parameter at redshift $z$ in a flat universe is
$H(z)=H_0[\Omega_m(1+z)^3 + \Omega_\Lambda]^{1/2}$, $dn/dM$ is the comoving
halo mass function from Jenkins (2001), and $\sigma_L$ is the cumulative
lensing cross--section in the lens plane for lensing amplification by a factor
$>\mu$ by a SIS (see Schneider, Ehlers \& Falco 1999). Note in particular that
the cross section scales with the magnification as $\sigma_L\propto \mu^{-2}$
for $\mu>2$, and as $\sigma_L\propto (\mu-1)^{-2}$ for $\mu<2$.

Fig.~\ref{fig:dtaudz} shows the a~priori probability $dp(\mu=2,z_s=5.99)/dz_L$
as a function of lens redshift that the $z=5.99$ quasar SDSSp
J130608.26+035636.3 (magnitude $z^*=19.47$) was magnified by at least a factor
$\mu_{\rm min}=2$, but has not produced two images separately detectable by
SDSS.  The middle curve in the figure shows the lensing probability for
multiple images with an image splitting of $\Delta\theta<1^{\prime\prime}$,
approximately the angular resolution of the SDSS (this corresponds to a
redshift--dependent upper--mass cutoff of $\sim 10^{12}~{\rm M_\odot}$ for the
lensing halos).  The bottom curve shows the probability for image splittings of
$\Delta\theta > 1^{\prime\prime}$, but subject to the constraint that the
fainter image must be below the SDSS detection limit of $z^\ast = 20.2$.  Note
that for an SIS lens, multiple (i.e., two) images are formed for $\mu\geq 2$,
with a flux ratio $\mu_1/\mu_2 = (\mu_{\rm tot} + 2) / (\mu_{\rm tot} -
2)$. Thus, the flux ratio is close to unity for high magnifications, and
increases monotonically for lower total magnifications.  In practice, the
constraint that only one image be detectable simply imposes a maximum total
magnification of $\mu_{\rm max}= 2 (10^{0.4\Delta m}+1)/(10^{0.4\Delta
m}-1)$. This source is $\Delta m =0.73$ mag above the detection threshold and
thus $\mu_{\rm max}\approx 6.172$.  The top curve in Fig.~1 shows the total
lensing probability.  As the figure shows, the total probability is very small,
of order 0.1 percent, and reaches a maximum near a lens redshift of $z_L=1.5$.

Fig.~\ref{fig:dtaudtheta} illustrates how the total lensing probability,
$p(\mu>2,z_s=5.99)$, depends on the maximum allowed image separation. For small
splitting angles the probability increases steeply, whereas for large angular
separations it flattens out. The dashed curve is for all halos, while the solid
curve only includes halos less massive than $10^{13}~{\rm M_\odot}$. As
discussed above, for the purposes of gravitational lensing, halos more massive
than $10^{13}~{\rm M_\odot}$ are better described by NFW profiles, which have a
much smaller cross-section for lensing than halos with SIS profiles. Including
the $M>10^{13}~{\rm M_\odot}$ halos using NFW profiles would result in a curve
that is nearly identical to the solid curve.  In both curves shown in this
figure, for splitting angles of $\Delta\theta>1^{\prime\prime}$, we require
that the fainter image must be below the SDSS detection threshold. As discussed
above, this excludes lensing events with $\mu\geq6.172$, and causes a
$\sim10\%$ decrease in the lensing probability for
$\Delta\theta>1^{\prime\prime}$. Note that according to
Figure~\ref{fig:dtaudtheta}, $\sim75\%$ of all lensing events have image
separations below $1''$, but only $\sim 10\%$ have image separations below
$0.1''$.

Our results in this section agree to within a factor of two with previous
estimates of the lensing probability and its dependence on lens redshift, e.g.\
by Kochanek (1998) and by Barkana \& Loeb (2000). We conclude that the a~priori
lensing probability for a $z \sim 6$ SDSS quasar along a random line of sight
is small, and dominated by lenses between redshifts $1\la z_L \la 2$.

\section{A~posteriori lensing probability}
\label{sec:aposteriori}
The a~priori lensing probability calculated in \S\ref{sec:apriori} reflects the
probability that lensing causes magnification by a factor of at least $\mu$
along a random line of sight. However, we are interested in the
{\it{a~posteriori}} probability that a given SDSS quasar has been lensed. Since
the SDSS can only detect $z\sim 6$ quasars that are at the bright-end tail of
the LF (if extrapolated from lower redshift), the magnification bias, defined
as the ratio between the a~posteriori and the a~priori probability, could well
be significant.

We determine the a~posteriori lensing probability from the observed quasar
LF. If we define $\Phi_{\rm int}(L)$ as the intrinsic space density (assuming
no lensing) of quasars with luminosity $L\pm dL/2$, then, given the intrinsic
differential lensing probability $dp/d\mu$, the space density of quasars that
will be observed at luminosity $L_{\rm obs} \pm dL_{\rm obs}/2$ and redshift
$z_{\rm s}$, and that are magnified by at least a factor $\mu_{\rm min}$ is
\begin{equation}
\Phi_{\rm obs}(L_{\rm obs},\mu_{\rm min})= \int_{\mu_{\rm min}}^\infty
d\mu \frac{dp}{d\mu} \Phi_{\rm int}(L_{\rm obs}/\mu) \frac{1}{\mu}.
\end{equation}

Similarly, the total space density of quasars observed at the same luminosity
and redshift, regardless of their amplifications, is $\Phi_{\rm obs}(L_{\rm
obs},0)$ and the a~posteriori probability that a quasar with observed
luminosity $L_{obs}$ is lensed by a minimum factor $\mu_{min}$ is
\begin{equation}
P(>\mu_{\rm min})=\frac{\Phi_{\rm obs}(L_{\rm obs},\mu_{\rm min})}
{\Phi_{\rm obs}(L_{\rm obs},0)}.
\end{equation}

The magnification bias and the a~posteriori lensing probability depend strongly
on the shape of the quasar LF. If the LF is steep or extends to faint
luminosities, the magnification bias can be strong.  Note that the
magnification bias depends only on the shape of the LF, not on its
normalization.  For the purpose of calculating $P(>\mu_{\rm min})$, we need, in
principle, to extend the intrinsic lensing probability distribution to cover
the full range of amplifications, including de--amplification, $0<\mu<2$.  We
shall use the following form of $dp/d\mu$:

\begin{equation}
\frac{dp}{d\mu} =
\left\{\matrix{
8\tau_2/\mu^3
 \hfill& (\mu\geq 2)  \hfill\cr
\tau_2/(\mu-1)^3
 \hfill& (\mu_0< \mu < 2)  \hfill\cr
\tau_2/(\mu_0-1)^3
 \hfill& (1\leq \mu < \mu_0)  \hfill\cr
0
\hfill& (\mu<1)  \hfill\cr} \right.
\label{eq:dpdmu}
\end{equation}

Here $\tau_2=\int_{2}^\infty d\mu (dp/d\mu)$ is the total optical depth for
forming double images (which depends on the source redshift, and is obtained
from equation~\ref{eq:lensdepth}).  The top two rows in
equation~(\ref{eq:dpdmu}) describe an SIS lens as usual, with a normalization
that ensures a smooth transition from the case when the source is doubly imaged
(but unresolved) to the singly imaged case.  In order to correctly model the
probability distribution near $\mu=1$, we would have to allow for
magnifications smaller than unity, which would require modeling the large-scale
density distribution of the universe.  Instead, we have imposed a critical
magnification $\mu_0\sim 1$, below which the probability is set to a constant.
The value of $\mu_0$ is chosen to be

\begin{equation}
\mu_0=1+\sqrt{\frac{3\tau_2}{2-\tau_2}},
\end{equation}
which enforces the normalization $\int_0^\infty d\mu\frac{dp}{d\mu} = 1$.  Note
that this simple characterization does not strictly conserve flux, since the
mean magnification $\int_0^\infty d\mu\frac{dp}{d\mu}\mu > 1$.  However, in
practice, we find that $\mu_0$ and the mean magnification are close to unity,
and our results are insensitive to our treatment of $dp/d\mu$ near $\mu=1$.

To calculate the a~posteriori lensing probability of the high redshift SDSS
quasars, we also need to know their intrinsic LF.  We here assume that the
intrinsic $z\sim 6$ quasar LF can be described by the double power law form
given in Pei (1995):
\begin{equation}
\Phi_{\rm int}(L)=\frac{\Phi_\ast/L_\ast}{
(L/L_\ast)^{\beta_l}+(L/L_\ast)^{\beta_h}}.
\end{equation}
The same form was used by Madau, Haardt, \& Rees (1999) and more recently by
Wyithe \& Loeb (2002), with different sets of parameters. The LF depends on
four parameters: the normalization $\Phi_\ast$, the faint-end slope $\beta_l$,
the bright-end slope $\beta_h$, and the characteristic luminosity $L_\ast$ at
which the LF steepens. The a~posteriori lensing probability is sensitive to the
last two parameters, $\beta_h$ and $L_\ast$ (and is strictly independent of the
normalization $\Phi_\ast$).

\begin{table}[t!]
\begin{center}
\caption{Parameters of the $z=6$ quasar luminosity functions.}
\label{tab:LF}
\begin{tabular}{lcccc}
\\
     &
\multicolumn{1}{c}{$\log(\frac{L_\ast}{{\rm L_{B,\odot}}})$}  &
\multicolumn{1}{c}{$\log(\frac{\Phi_\ast}{{\rm L^{-1}_{B,\odot}}{\rm Gpc^{-1}}})$} &
\multicolumn{1}{c}{$\beta_l$} &
\multicolumn{1}{c}{$\beta_h$} \\
\tableline
\tableline
\multicolumn{5}{c}{Varying Break} \\
\tableline
Pei (1995)    &  10.10  &  6.57  &  1.64  &  3.52 \\
Pei (1995)$^2$&  10.87  &  4.61  &  1.64  &  3.52 \\
WL (2002)     &  11.95  &  1.91  &  1.64  &  3.52 \\
MHR (1999)    &  11.55  &  2.76  &  1.64  &  3.43 \\
\tableline
\multicolumn{5}{c}{Varying Bright End Slope} \\
\tableline
steep         &  11.55  &  4.63  &  1.64  &  4.5 \\
medium        &  11.55  &  2.88  &  1.64  &  3.5 \\
shallow       &  11.55  &  1.14  &  1.64  &  2.5 \\
\end{tabular}
\end{center}
\end{table}

\begin{figure}[ht!]
\begin{center}
\plotone{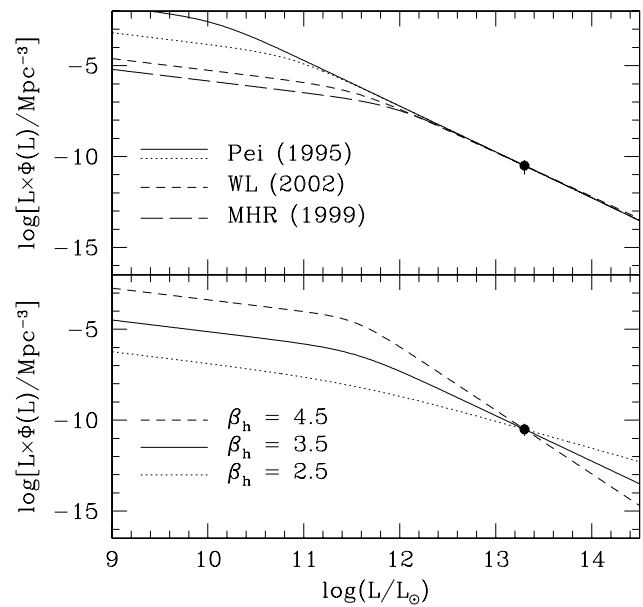}
\caption{The seven different LFs used to compute lensing probabilities. The LFs
are all described by a double power law with the same faint--end slope of
$\beta_l=1.64$, and are shown at redshift $z=6$. The data point is from Fan et
al.\ (2001) and represents the space density of luminous quasars at $z\approx
6$, inferred from the discovery of the four SDSS quasars ($\log(L\Phi/{\rm
Mpc}^{-3}) = -10.5$ at $\log (L_B/{\rm L_{B,\odot}}) = 13.3$). The {\it upper
panel} demonstrates the effect of varying the break of the LF. From top to
bottom the four LFs have $\log (L_\ast/{\rm L_{B,\odot}}) \approx 10.1$, 10.9,
11.5, and 11.9 respectively. Apart from the normalization $\Phi_\ast$, which
has been adjusted to ensure consistency with the SDSS data point, the curves
correspond to the model of Pei (1995) (solid), the Pei (1995) model with a 6
times larger $L_\ast$ (dotted), the model of Wyithe \& Loeb (2002) (short
dashed), and Madau et al.\ (1999) model (long dashed). The Wyithe \& Loeb
(2002) model (short dashed) has $\beta_h = 3.43$, while the other models have
$\beta_h = 3.52$. The {\it lower panel} demonstrates the effect of varying the
bright--end slope $\beta_h$ from 2.5 to 4.5. All three models have $\log
(L_\ast/{\rm L_{B,\odot}}) \approx 11.5$.  See Table~\ref{tab:LF} for a summary
of the LF parameters.}
\label{fig:lf}
\end{center}
\end{figure}

Through different combinations of slopes and breaks, we present seven possible
LFs at redshift $z=6$, which are illustrated in Fig.~\ref{fig:lf}. All models
have $\beta_l = 1.64$ (Pei 1995) and the normalization $\Phi_\ast$ has been
adjusted so that each curve passes through the Fan et al.\ (2001b) data point
corresponding to the space density implied by the four $z \sim 6$ quasars. The
curves in the upper panel all have $\beta_h \approx 3.5$, but $L_\ast$ varies
from $\approx 10^{10}~L_{B,\odot}$ for the Pei (1995) model (solid curve) to
$\approx 10^{12}~L_{B,\odot}$ for the model of Madau et al.\ (1999). The curves
in the bottom panel all have $L_\ast\approx 10^{11.5}~L_{B,\odot}$, but have
different values of $\beta_h$.  The parameters of the LFs are summarized in
Table~\ref{tab:LF}.  The values are given at redshift $z=6$; we allow for
slightly different values for the characteristic luminosity over the range
$5.8<z<6.28$ by including the redshift dependence of $L_\ast$ as parameterized
in each case by Pei (1995), Madau et al.\ (1999), and Wyithe \& Loeb (2002).

The lensing probability depends on the intrinsic LF, which is not directly
observable. In using the Fan et al.\ (2001b) data point as a constraint on the
intrinsic LF, we thus implicitly assumed that lensing does not have a large
overall effect on the LF. If magnification bias is significant, then this
assumption breaks down for the bright-end of the LF. Thus, to match the
observations, the normalizations of the intrinsic LFs should in fact be lowered
by a factor equal to the magnification bias for the observed quasars. However,
for the purpose of computing the a~posteriori lensing probability this is
irrelevant, since the magnification bias is independent of the normalization of
the intrinsic LF.

\begin{figure}[ht!]
\begin{center}
\plotone{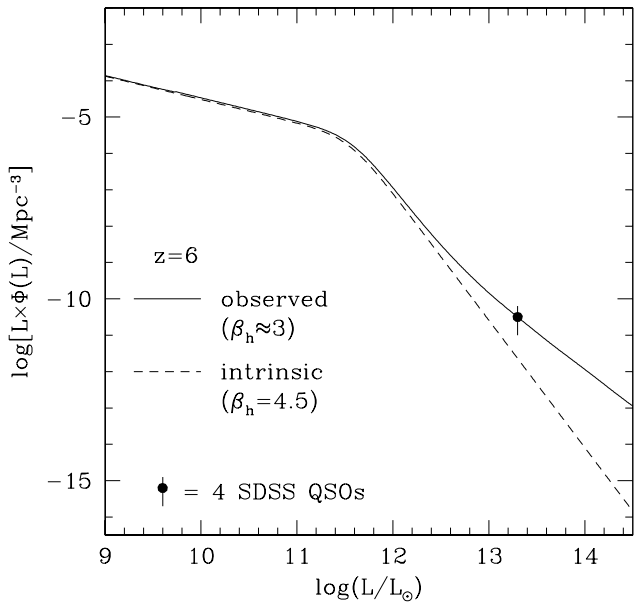}
\caption{A comparison of the intrinsic and observed LFs for our most extreme
model, which has an intrinsic bright-end slope of $\beta_h=4.5$ (see dashed
curve in lower panel of Fig.~\ref{fig:lf}).  The lensing probability at $z=6$
in this case is high, and the apparent LF is increased by over an order of
magnitude at the luminosity of the SDSS quasars. The apparent slope is
flattened to $\beta\approx 3$, and the model is therefore consistent with the
upper limit on slope of the apparent LF from the SDSS ($\beta<4.3$ at 99
percent confidence, Fan et al. 2001).}
\label{fig:lfobs}
\end{center}
\end{figure}

The SDSS observations provide an upper limit on the slope of the bright-end of
the observed LF: $\beta_{h,obs} < 4.3$ at 99 percent confidence (Fan et al.\
2001a,b). If magnification bias is important, then this constraint will always
be satisfied since lensing by an isothermal sphere will result in a bright-end
slope of $\beta_{h,obs} = 3.0$ (the intrinsic differential lensing probability
$dp/d\mu \propto \mu^{-3}$ for $\mu > 2$). This is illustrated in
Fig.~\ref{fig:lfobs}, which shows the intrinsic and observed LFs for our most
extreme model, which has $\beta_h = 4.5$ (dashed curve in the lower panel of
Fig.~\ref{fig:lf}). Since the magnification bias increases with $\beta_h$,
observations cannot directly constrain the steepness of the intrinsic,
bright-end slope of the LF. However, we will see below that the slope can be
constrained indirectly, by measuring the fraction of multiply imaged quasars.

\begin{figure}[ht!]
\begin{center}
\plotone{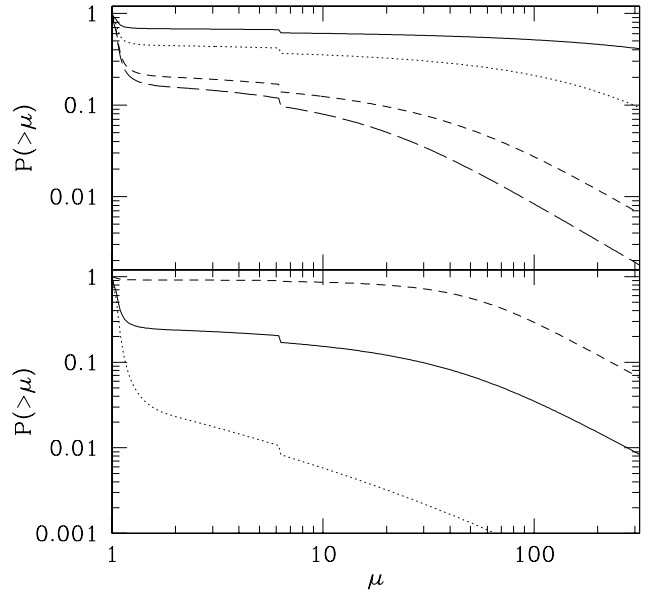}
\caption{Lensing probability as a function of minimum magnification $\mu$ for
the $z=5.99$ SDSS quasar and the seven different LFs shown in
Fig.~\ref{fig:lf}. Note that the total lensing probability is sensitive to
variations of both the slope $\beta_h$ and break $L_\ast$, and can be high.
For $\mu>6.172$, both lensed images would appear brighter than SDSS's
threshold, and we impose the condition $\Delta\theta<1^{\prime\prime}$ on their
splitting angle to avoid detection by SDSS. On the other hand, the fainter
image is too faint to be detectable, and we allow lensing with arbitrarily
large splitting angles for $\mu<6.172$.}
\label{fig:lensprob}
\end{center}
\end{figure}

\begin{deluxetable}{l c c c c}
\tablecaption{Parameters and lensing probabilities of the four $z\approx 6$ 
SDSS quasars\tablenotemark{a}\label{tab:qsos}}
\tablecolumns{5} 
\tablewidth{0pc} 
\tabletypesize{\scriptsize}
\tablehead{
\colhead{} &
\colhead{1044-0125} &
\colhead{0836-0054} &
\colhead{1306-0356} &
\colhead{1030-0524} \\}
\startdata 
Redshift                       &  5.80     &  5.82    &  5.99    &  6.28    \\
Magnitude [$z^\ast$]           &  19.23    &  18.74   &  19.47   &  20.05   \\
\tableline
$\tau_2$  [any $\Delta\theta$] &  0.00314  &  0.00315 &  0.00326 &  0.00344 \\
$\tau_2$  [$\Delta\theta<1''$] &  0.00238  &  0.00239 &  0.00247 &  0.00260 \\
$\mu_{\rm min}$                &  4.771    &  3.410   &  6.172   &  28.99   \\
$P(\mu>\mu_{\rm min})$ [$\Delta\theta>1'', 
                 \beta_h=4.5$] &  0.220    &  0.231   &  0.218   &  0.150 \\
$P(\mu>10)$\,\,\,\,\,\, [$\Delta\theta<1'', 
                 \beta_h=4.5$] &  0.877    &  0.939   &  0.862   &  0.788 \\
\enddata 
\tablenotetext{a}{The first two lines list the redshift and magnitude of each
source. The 3$^{\rm rd}$ and 4$^{\rm th}$ lines give the optical depth to
lensing with amplification $\mu\geq 2$ with and without including events whose
splitting angles are larger than $1''$.  The 5$^{\rm th}$ line lists the minimum
total magnification for which both images are above the SDSS magnitude
threshold of $z^\ast=20.2$.  The last two lines list the lensing probabilities,
with magnification bias taken into account, in our most extreme model with a
bright--end quasar LF slope of $\beta_h=4.5$.  The 6$^{\rm th}$ line gives the
probability for a event that produces two separately detectable images for
SDSS.  The last line gives the probability for an unresolved lensing event with
total magnification $\mu\geq 10$.}
\end{deluxetable}

Fig.~\ref{fig:lensprob} shows the a~posteriori probabilities for the $z=5.99$
SDSS quasar to be magnified by at least a factor $\mu$, subject to the
constraint that only one image is detectable by SDSS.  The small discontinuity
in the probabilities at $\mu=6.172$ occurs because for $\mu<6.172$ the fainter
image is too faint to be detectable, and we do not impose any constraint on the
splitting angle.  On the other hand, for $\mu>6.172$ both images would be
brighter than SDSS's threshold, and we therefore impose the condition
$\Delta\theta<1^{\prime\prime}$ on the splitting angle.  The seven curves shown
in Fig.~\ref{fig:lensprob} correspond to the seven intrinsic LFs in
Fig.~\ref{fig:lf}. From the figure it is clear that a wide range of
probabilities is possible for a given minimum magnification. In particular,
because of the lack of constraints on the intrinsic LF, they can be
significantly higher than the maximum value of $\sim 30\%$ found by Wyithe \&
Loeb (2002), who considered a more restricted range of LFs (the short dashed
curve in the upper panel in Fig.~\ref{fig:lf} represents their most extreme
model that results in their maximum lensing probabilities). For example, if
$L_\ast$ is small, as for the Pei (1995) model (solid curve in upper panel),
then the probability that the quasar has been magnified by more than a factor
of ten is about sixty percent. The lensing probability also becomes very large
if the bright-end slope is steep: $P(\mu > 10)$ is close to unity if
$\beta_h=4.5$ (dashed curve in the lower panel).  In Table~\ref{tab:qsos}, in
the last line, we list the lensing probability for each of the four quasars in
this model.

\vspace{\baselineskip}

\section{Constraining the shape of the luminosity function}
\label{sec:LF}

In the previous section we have demonstrated that it is possible that most of
the $z\sim 6$ SDSS quasars are strongly magnified by gravitational lensing, and
that the expected lensing probability depends very strongly on the shape of the
intrinsic quasar LF.  Because of the unknown magnification bias, surveys that
probe only the bright-end of the LF ($L>L_\ast$), cannot directly measure the
intrinsic LF. However, it is possible to constrain the intrinsic LF, in
particular the characteristic luminosity $L_\ast$ and the bright-end slope
$\beta_h$, by measuring the fraction of multiply imaged quasars down to a fixed
limiting angular resolution.

The SDSS survey has so far discovered four $z\sim 6$ quasars (Fan et al.\
2001), all of which appear to be single images at the angular resolution ($\sim
1 ''$) and detection limit ($z^\ast = 20.2$) of the survey.  The parameters of
these quasars are listed in Table~\ref{tab:qsos}.  Lensing by an isothermal
sphere always yields a double image if the total magnification is greater than
two. Thus, if these object are strongly lensed, then the extra image must have
escaped attention either because the splitting angle was too small for SDSS to
resolve the images or because one of the images was too faint to be detected.
In the previous section, when we computed the a~posteriori probabilities for
lensing, we took these limitations of the survey into account.

In this section, we relax these limitations, and compute the complementary
quantity: the probability of lensing events that would produce multiple images,
both of which would have been detectable by SDSS (i.e., we require that both
images be above the detection threshold, and separated by $>1 ''$). The lack of
such a detection can be used to constrain models for the LF.  In particular,
for a given quasar LF, we can compute the likelihood that {\it at least one} of
the four SDSS quasars would exhibit two images. This is given by
$P=1-(1-P_1)\times(1-P_2)\times(1-P_3)\times(1-P_4)$, where $P_i$ is the
probability for the $i^{\rm th}$ quasar to be doubly imaged.  We consider
variations in the two parameters $L_\ast$ and $\beta_h$ of the quasar LF, and
compute $P$ as a function of $L_\ast$ and $\beta_h$.  As an example, in
Table~\ref{tab:qsos}, in the 6$^{\rm th}$ line, we list the probability for
each quasar to be doubly imaged in our most extreme model for the quasar LF.

\begin{figure}[ht!]
\begin{center}
\plotone{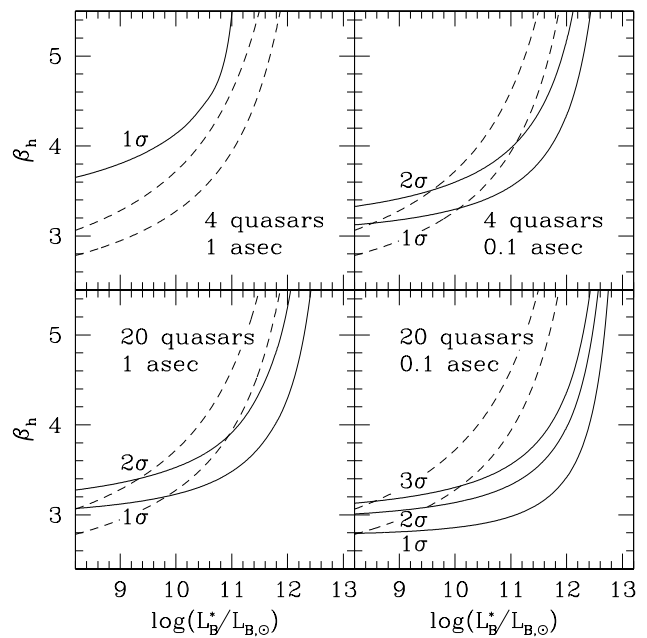}
\caption{Constraints on the parameters of the redshift $z=6$ quasar LF from the
non--detection of gravitational lenses.  The solid curves show likelihood
contours for ruling out a given LF, if no gravitational lenses are found.  The
upper left panel shows the $1 \sigma$ constraint from the existing four SDSS
quasars, assuming that double images are resolvable if they are separated by
$>1 ''$, and are both above the magnitude threshold of $z^*=20.2$. The
constraint is weak.  The other three panels demonstrate how the constraint can
be improved if no lenses are found despite (i) increasing the angular to $0.1
''$ (upper right panel); (ii) increasing the sample size to a total of 20
quasars (lower left panel); or (iii) both (lower right panel).  The $3\sigma$
contour is off the scale of the plots, except in the lower right panel; the
$2\sigma$ contour is off the scale in the upper left panel. The dashed curves
show contours of fixed ionizing background $\Gamma_{-12}=0.1$ and
$\Gamma_{-12}=2$.}
\label{fig:contour0}
\end{center}
\end{figure}

In the upper left panel of Fig.~\ref{fig:contour0}, we show the constraint on
the intrinsic LF resulting from the non--detection of any multiple images in
the four quasars reported in Fan et al.\ (2001). The solid curve corresponds to
the constant $P=0.68$, i.e., models that lie above the solid curve are excluded
at the 1~$\sigma$ level. The analogous 2 and 3 $\sigma $ contours are off the
scale of the plot. The SDSS observations do not yet place strong constraints on
the intrinsic LF. This can be understood as follows: the intrinsic probability
for lensing with multiple images to $z=6$ when all lensing halos are considered
(up to a mass $10^{13}~{\rm M_\odot}$, see \S~\ref{sec:apriori}) is
$\tau_2=0.00326$, but when splittings of $> 1''$ are excluded, the probability
is $\tau_2=0.00247$.  In other words, only $\sim 25\%$ of all lensing events
would be resolved at $> 1''$. Even for an arbitrarily large magnification bias,
SDSS would only have a $\sim 25\%$ chance of resolving the multiple images of a
given quasar (see Table~\ref{tab:qsos}).  Note that the $z=6.28$ source is very
close to the SDSS magnitude threshold, and the fainter image would escape
detection unless the total magnification is $\mu>29$. As a result, the
probability for producing an additional detectable image for this source is
smaller than for the other three quasars.

\begin{figure}[ht!]
\begin{center}
\plotone{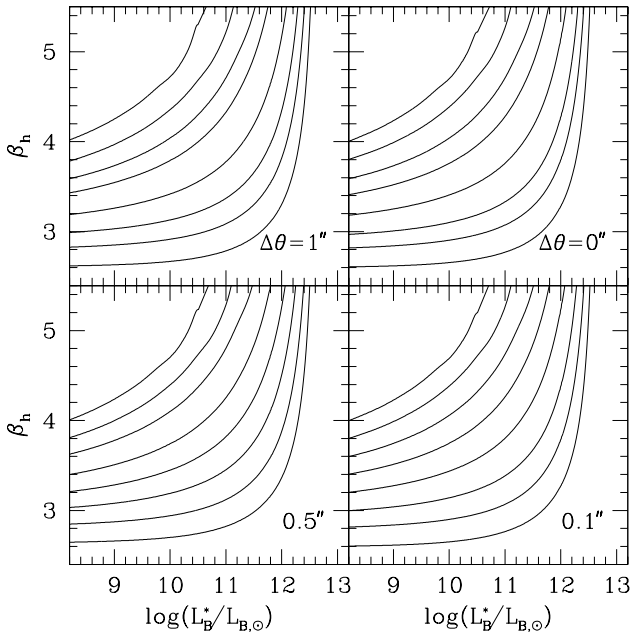}
\caption{Likelihood contours as in Fig.~\ref{fig:contour0}, but for a single
$z=6$, $z^*=20$ quasar.  The curves in each of the four panels show constant
probabilities for a lensing event that is detectable at different angular
resolutions (the upper right panel assumes infinite resolution, i.e., it
includes all multiple lensing events).  Each panel shows 8 curves,
corresponding to the following probabilities (bottom to top, in percent): (i)
Upper left panel: 1, 2, 4, 10, 20, 23, 24, 24.18; (i) Upper right panel: 4, 8,
15, 40, 80, 95, 99.3, 99.9; (i) Lower left panel: 2, 4, 9, 20, 35, 43, 44.3,
44.55; (i) Lower right panel: 3, 6, 13, 33, 60, 72.5, 75.4, 75.86.  Note the
``saturation'' of the probabilities near the top contour in each panel, caused
by the finite fraction of lensing events that produce splittings above a given
angular resolution.}
\label{fig:contour}
\end{center}
\end{figure}

In the remaining three panels of Fig.~\ref{fig:contour} we illustrate how future
observations would constrain the LF if no multiple images were detected. The
bottom left panel shows the results for 20 objects with properties
identical\footnote{In increasing the sample size, we simply replicated the
current sample of quasars five times. Since the lensing probabilities are
similar for quasars near the detection threshold and near $z=6$, our results
are not sensitive to this particular choice.} to the four already discovered
quasars.  If SDSS were to discover 20 $z\sim 6$ quasars, and none of them
showed a second image, then the more extreme intrinsic LFs plotted in
Fig.~\ref{fig:lf} (the Pei 1995 and the $\beta_h = 4.5$ models) would be ruled
out at the $2~\sigma$ level.

The upper right panel shows that nearly identical constraints can be obtained
by observing the four already known quasars with an angular resolution of $0.1
''$, a resolution typical for observations with the \emph{Hubble Space
Telescope} (for the cases of a maximum splitting angle of $0.1 ''$ we assumed
that sufficient depth is reached that both images are detectable, and did not
impose a magnitude cut, or a constraint on the flux ratio). Finally, the bottom
right panel shows that observations of 20 quasars at $0.1 ''$ resolution would
enable us to put very strong constraints on the intrinsic LF, with the current
``ball-park'' LFs, obtained by extrapolations from lower redshifts, ruled out
at the $\sim 3~\sigma$ level.

In these examples we assumed that no multiple images are detected. Similar
calculations can of course be done if future observations would detect one or
more multiple images. In that case the likelihood contours can be closed and
the parameters of the LF would be constrained both from above and from below.
To enable interpretation of any outcome of a future search for lenses,
Figure~\ref{fig:contour} shows probability contours for multiple lensing for a
single $z=6$, $z^*=20$ quasar.  The curves in each of the four panels show
constant probabilities for a lensing event detectable at four different angular
resolutions (the upper right panel assumes infinite resolution, i.e., it
includes all multiple lensing events).  Each panel shows 8 curves,
corresponding to the following probabilities (bottom to top, in percent): (i)
Upper left panel: 1, 2, 4, 10, 20, 23, 24, 24.18; (i) Upper right panel: 4, 8,
15, 40, 80, 95, 99.3, 99.9; (i) Lower left panel: 2, 4, 9, 20, 35, 43, 44.3,
44.55; (i) Lower right panel: 3, 6, 13, 33, 60, 72.5, 75.4, 75.86.  Note the
``saturation'' of the probabilities near the top contour in each panel, caused
by the finite fraction of lensing events that produce splittings above a given
angular resolution.

Varying the intrinsic LF does not only affect the lensing probability, but also
changes the total amount of emitted quasar light, including the ionizing
emissivity.  Since the ionizing background can in principle be measured, this
can provide an additional constraint on the quasar LF. Observations of the mean
absorption in the Lyman-$\alpha$ forest have been used recently to infer the
ionizing background at $z\sim 6$ (Becker et al. 2001; Cen \& McDonald 2001;
McDonald \& Miralda-Escud\'e 2001; Fan et al. 2002), yielding ionization rates
per hydrogen atom of $\Gamma_{-12} \equiv \Gamma / 10^{-12} {\rm s}^{-1}
\approx 0.1$.  The dashed curves in Fig.~\ref{fig:contour0} show contours of
fixed quasar emissivity, corresponding to $\Gamma_{-12} = 0.1$ and 2
respectively.  Our method for computing the ionization rate from the LF is
described in more detail in the Appendix below. At face value, LFs that lie
above the lower dashed curve over--produce the ionizing background; those above
the upper dashed curve by a factor of $>20$. However, in practice this result
has large uncertainties.  First, although the adopted quasar LF yields the
total ionizing emissivity, the prediction of the ionizing background involves
the mean free path of ionizing photons, which is highly uncertain at the
redshifts of interest here. A small mean free path would reduce the background.
Second, a smaller fraction of ionizing photons might escape from the dense
surroundings of high--redshift quasars than from their lower--redshift
counterparts.

\section{Discussion and Conclusions}

In this paper, we computed the expected lensing probabilities of $z\sim6$
quasars by intervening galaxies. The a~posteriori probability that a $z \sim 6$
quasar is strongly lensed (by a factor $\mu>10$ in flux enhancement), but
without producing two images detectable in the Sloan Survey, can be very large,
depending on the shape of the LF. The LFs we use are consistent with other
available observational constraints, yet these constraints still allow for a
wide range of possibilities.

Though an {\it observed} bright-end slope of $\beta_h=4.3$ could be ruled out
at the 99 percent confidence level (Fan et al.\ (2001)), lensing alters the
slope so that the apparent LF will have a slope close to $\sim 3.0$ if
magnification bias is important. As illustrated in Fig.~\ref{fig:lfobs}, even
the steepest intrinsic LF we consider, with a bright end slope of $4.5$, is
consistent with the SDSS limit on slope. This will be true for arbitrarily
steep slopes, provided that the characteristic luminosity is not increased.

In a similar probability calculation, Wyithe \& Loeb (2002) concluded that the
observed probability reaches $30 \%$ that a single $z \sim 6$ quasar is lensed
and magnified by a factor of 10 or more. If we restrict our analysis to the
choices of LFs considered in their paper, we agree with this result. However,
we here use a wider range of quasar LFs and conclude that the lensing
probability can reach essentially 100\%.  As a result, the current observations
are consistent with all four $z \sim 6$ SDSS quasars being strongly lensed.

If the probability that all four quasars are lensed were high, then this would
alleviate the problematic time constraints on assembling supermassive black
holes at the earliest stages in the evolution of the universe . However, in a
separate analysis, based on the apparently large size of the ionized region
around the SDSS quasar at $z=6.28$, Haiman \& Cen (2002) place a strong
constraint on the lensing magnification of this source, and find $\mu<5$. This
result depends primarily on the assumption that the source is embedded in a
neutral (rather than ionized) intergalactic medium.

We have also considered the probabilities for lensing events that would have
been detectable by SDSS. The lack of such detections can be used to place
constraints on the quasar LF.  Although constraints from the current four
quasars are mild, the situation is likely to improve.  As illustrated in
Fig.~\ref{fig:contour0}, increasing the SDSS sample from 4 to 20 objects,
something that is likely to happen over the next few years, would allow one to
rule out interesting quasar LF models. The expected probability of detecting
multiple images is also sensitive to the angular resolution of the
observations. In terms of the constraints on the shape of the intrinsic LF,
observing four quasars at a resolution of $0.1 ''$ is roughly equivalent to
observing 20 objects at the resolution of the SDSS ($1 ''$).

Hence, upcoming observations, in particular with the \emph{Hubble Space
Telescope}, should reveal whether a significant fraction of the $z\sim 6$
quasars is lensed, and will allow us to place strong constraints on their
intrinsic LF.

\acknowledgements

We thank Michael Strauss and Xiaohui Fan for many useful discussions.  The work
presented here is based on the senior thesis of JC at Princeton University. She
thanks Neta Bahcall for her encouragement and advice.  This work was supported
by NASA through the Hubble Fellowship grant HF-01119.01-99A, awarded to ZH by
the Space Telescope Science Institute, which is operated by the Association of
Universities for Research in Astronomy, Inc., for NASA under contract NAS
5-26555. JS acknowledges support from the W.~M.~Keck foundation.

\vspace{\baselineskip}

\section*{Appendix}
Here we discuss constraints on the high--redshift quasar LF from their
integrated ionizing radiation. Given an LF, we compute the total comoving
emissivity as $\nu \epsilon_\nu=\int L \Phi(L) dL$.  This gives $\nu
\epsilon_\nu$ at rest--frame 4400\,\AA; however, with a spectral slope of
$\epsilon_\nu\propto \nu^{-1}$ this is independent of wavelength, and has the
same value at 912\,\AA.  The background ionization rate per hydrogen atom
$\Gamma$ is then given by
\begin{equation}
\Gamma=\frac{\pi\sigma_H}{h_{\rm P}}\lambda_{\rm mfp}\frac{(1+z)^3}{4\pi}
\frac{\nu\epsilon_\nu}{\nu_{\rm H}}~~~~~{\rm s^{-1}}, 
\end{equation}
where we have assumed $\epsilon_\nu\propto\nu^{-1}$, $\sigma_H=6.3\times
10^{-18}~{\rm cm^{-2}}$ is the hydrogen ionization cross section at the Lyman
limit (and we assumed $\sigma_H(\nu)\propto \nu^{-3}$ above the threshold),
$h_{\rm P}$ is the Planck constant, $\nu_{\rm H}=3.29\times 10^{15}$Hz is the
frequency of a Lyman limit photon, and $\lambda_{\rm mfp}$ is the mean free
path of ionizing photons.  We also define $\Gamma_{-12}=\Gamma/10^{-12}~{\rm
s^{-1}}$.  We assume that the effective mean free path corresponds to a
redshift interval of $\Delta z=0.17$ at $z=3$ (see Haardt \& Madau 1996 and
Steidel, Pettini \& Adelberger 2001), and scales with redshift as $\lambda_{\rm
mfp}\propto (1+z)^{-6}$ towards higher redshift (Cen \& McDonald 2002).  Based
on recent measurements (e.g. Cen \& McDonald 2002, see also Becker et al. 2001;
McDonald \& Miralda-Escud\'e 2001 and Fan et al. 2002), we require
$\Gamma_{-12}<0.1~{\rm s^{-1}}$ at z=6.  We find that in some of our models, at
$z=6$, this bound is violated by up to a factor of 13: $\Gamma_{-12}=1.3$ (Pei
LF); $\Gamma_{-12}=0.39$ (WL LF with a bright--end slope of $\beta_h=4.5$);
$\Gamma_{-12}=0.08$ (Pei LF with 6 times higher $L_\ast$).  More generally,
contours of constant $\Gamma_{-12}=0.1$ and $\Gamma_{-12}=2$ are shown in
Figure~\ref{fig:contour0}.

The apparently large ionizing emissivities imply that in our LF with the lowest
$L_\ast$, the background is over--produced, although this result is subject to
several uncertainties : (1) the escape fraction of ionizing photons may be only
$\sim 10\%$ at $z=6$; (2) the mean free path evolves more steeply than
$(1+z)^{-6}$; (3) the faint--end slope is considerably shallower than
$\beta_l=1.64$, or (4) there can be significant systematic errors on the
observational determination of $\Gamma_{-12}$, since it relies on cosmological
simulations for modeling the temperature and density field.

\end{document}